\documentclass[runningheads]{llncs} 
\usepackage[T1]{fontenc}
\usepackage{graphicx}
\usepackage{booktabs}   

\usepackage{placeins}
\begin{document}
\title{Behind Python: The Languages That Power AI}
\titlerunning{Behind Python: The Languages That Power AI}
%

\author{Juan P. Licona-Luque \and
Beatriz A. Bosques-Palomo \and
Nezih Nieto-Guti\'errez \and
Gustavo de los R\'ios-Alatorre \and
Luis A. Mu\~noz-Ubando}
\authorrunning{J. P. Licona-Luque et al.}
%
\institute{Tecnol\'ogico de Monterrey, Monterrey, Nuevo Le\'on, Mexico\\
\email{\{juan.licona, bosques.beatriz\}@exatec.tec.mx}\\
\email{\{nezihniegu, amunoz, A01410922\}@tec.mx}}
\maketitle              
\begin{abstract}
Python dominates AI development, yet the numerical work behind
frameworks like PyTorch and NumPy is executed in C, C++, or Rust.
When a developer must implement an algorithm without such
libraries---because none exists, the target is resource-constrained,
or a new system is being built---which language should they choose?
This paper answers that question empirically. Five algorithms covering
data mining ($k$-means), machine learning ($k$-NN), neural networks
(MLP with backpropagation), computational intelligence (genetic
algorithm), and fuzzy systems (Mamdani inference) are implemented from
scratch in Python, C, C++, Rust, Go, and Julia. All implementations
share a common pseudo-random generator, consume identical inputs, and
produce bit-identical outputs, so every measured difference reflects
the language rather than the computation. Three performance tiers
emerge: C and C++ are effectively tied; Rust trails them by $9\%$
(geometric mean); Julia runs $3.3\times$ slower than C and Go
$5.0\times$; Python sits at $315\times$. Memory tells a different
story---Julia's JIT runtime carries a fixed ${\sim}\,224$\,MiB
footprint regardless of workload, while C, C++, and Rust stay below
$6$\,MiB. Crucially, rankings are not stable: Go's slowdown swings
from $2.6\times$ on $k$-NN to $8.0\times$ on $k$-means, showing that
workload characteristics can shift a language's position by a full
tier. The results provide concrete, per-workload guidance for choosing
an implementation language in AI systems.

\end{abstract}
\section{Introduction}

Artificial intelligence (AI) systems are commonly developed using a
layered software architecture. At the application level, data scientists
and engineers rely heavily on high-productivity languages such as Python
for experimentation, model development, and orchestration. However, the
computational kernels that enable modern AI frameworks to achieve
competitive performance are frequently implemented in lower-level
compiled languages. NumPy relies extensively on C-based numerical
routines, while frameworks such as PyTorch and TensorFlow build their
performance-critical components in C++ and CUDA. More recently, Rust has
gained adoption in AI infrastructure projects such as Hugging Face
Tokenizers and Polars due to its combination of performance and memory
safety. Julia, in contrast, was designed specifically to bridge the
traditional gap between productivity and high-performance numerical
computing through just-in-time compilation and a unified language
ecosystem~\cite{bezanson2017,numpy2020,paszke2019pytorch,churavy2022}.

This reality raises a practical question that remains insufficiently
explored in literature. Developers often need to implement algorithms
themselves, whether because no suitable library exists, a target
environment imposes strict resource constraints, or they are building the
next generation of AI infrastructure. In these situations, which
programming language offers the best balance between execution speed,
memory efficiency, and development effort? While compiled languages are
often assumed to offer better performance than interpreted languages, the
magnitude of this advantage and the extent to which it depends on the
characteristics of specific AI workloads remain unclear.

Prior research has extensively compared programming languages from the
perspectives of execution time, energy consumption, memory usage, and
programmer productivity. Studies such as Prechelt's empirical comparison
of seven languages~\cite{prechelt2000}, the Computer Language Benchmarks
Game~\cite{clbg}, and the large-scale evaluation of energy efficiency by
Pereira et al.~\cite{pereira2017} provide valuable insights into language
design trade-offs. Similarly, Bugden and Alahmar investigated Rust's
ability to deliver performance comparable to C and C++ while maintaining
stronger safety guarantees~\cite{bugden2022}. However, these studies
largely rely on synthetic workloads or systems-oriented benchmarks such as
recursive algorithms, sorting routines, numerical kernels, and
tree-processing tasks. Although useful for evaluating language runtimes
and compilers, these workloads do not necessarily reflect the
computational patterns found in artificial intelligence algorithms.

At the opposite end of the spectrum, many comparisons of programming
languages in AI contexts evaluate implementations built on optimized
libraries and frameworks. In such cases, performance measurements
primarily reflect the underlying library implementations rather than the
host language itself. For example, comparing Python and Julia
implementations that both invoke highly optimized native kernels reveals
relatively little about the intrinsic costs and benefits of the
languages. Consequently, there remains a gap between traditional
programming-language benchmarks and real-world AI algorithm
implementations.

This paper addresses that gap through a controlled empirical study of six
programming languages: Python, C, C++, Rust, Go, and Julia. These
languages were selected because of their significant roles in modern AI
infrastructure, with Python and Julia emphasizing developer productivity,
C and C++ serving as the dominant implementation languages for
performance-critical kernels, Rust emerging as a memory-safe alternative
adopted in projects such as Hugging Face Tokenizers and Polars, and Go
playing an important role in model-serving and cloud-native AI systems.
They are evaluated using five representative algorithms spanning major
branches of artificial intelligence: $k$-means clustering, $k$-nearest-neighbors
classification, multilayer perceptron training with backpropagation,
genetic algorithms, and Mamdani fuzzy inference systems. Each algorithm is
implemented from scratch without external numerical or machine-learning
libraries, ensuring that observed differences arise from the languages,
compilers, and runtimes themselves rather than third-party optimizations.

To guarantee fairness and reproducibility, all implementations share
identical algorithmic specifications, deterministic input generation, and
a common pseudo-random number generator. Furthermore, every implementation
is validated through cross-language checksum verification to ensure that
identical computations are performed across all platforms. Each language
is evaluated along five dimensions: execution time, peak memory
consumption, binary size, source-code size, and compilation time.

The study is organized around three research questions:
\begin{description}
  \item[RQ1] How do Python, C, C++, Rust, Go, and Julia compare in
    execution time and memory consumption when implementing AI algorithms
    from scratch?
  \item[RQ2] What productivity--performance trade-offs emerge when runtime
    metrics are considered alongside implementation complexity and
    compilation cost?
  \item[RQ3] Are language rankings consistent across different categories
    of AI algorithms, or do workload characteristics significantly
    influence the outcome?
\end{description}

This work makes three main contributions. First, it provides a controlled
benchmark of five AI algorithms implemented from scratch in six
programming languages. Second, it establishes a reproducible methodology
that ensures fair cross-language comparisons by eliminating differences in
algorithm implementation and output. Third, it quantifies the trade-offs
between execution speed, memory consumption, compilation cost, and
implementation effort, providing practical insights for language
selection in AI development.

\section{Related Work}\label{sec:related}

\subsection{Programming-Language Performance Evaluation}
The empirical comparison of programming languages has been an active
research area for decades. One of the earliest influential studies was
conducted by Prechelt~\cite{prechelt2000}, who compared seven programming
languages using a common software-development task and highlighted the
trade-offs between development effort and execution performance. Since
then, the Computer Language Benchmarks Game~\cite{clbg} has become one of
the most widely referenced resources for cross-language performance
comparisons, providing implementations of standardized computational
benchmarks across dozens of programming languages. Nanz and Furia further
expanded this line of research by analyzing implementations from the
Rosetta Code repository and comparing languages in terms of performance,
conciseness, and other software quality attributes~\cite{nanz2015}.

More recent work has expanded the scope of language evaluation beyond
execution speed. Pereira et al.~\cite{pereira2017} analyzed twenty-seven
programming languages and demonstrated significant relationships among
execution time, memory consumption, and energy efficiency. Their results
reinforced the notion that language choice can have measurable
consequences not only for performance but also for resource utilization.
Similarly, Bugden and Alahmar~\cite{bugden2022} examined Rust's
safety-oriented design and found that it can achieve performance
comparable to traditional systems languages while reducing classes of
memory-related errors.

Although these studies provide valuable evidence regarding language
behavior, most rely on synthetic or systems-oriented workloads such as
recursive computations, sorting algorithms, regular-expression processing,
and numerical kernels. Consequently, their findings cannot be directly
generalized to AI workloads, whose computational characteristics often
include iterative optimization, dense numerical computation,
population-based search, and rule-based inference. The present work adopts
the methodological strengths of this literature while focusing specifically
on algorithms representative of artificial intelligence.

\subsection{Programming Languages in Modern AI Infrastructure}
Modern AI software is characterized by a separation between user-facing
productivity layers and performance-critical computational kernels. Python
dominates AI development due to its extensive ecosystem and ease of use,
yet the underlying numerical computations are frequently delegated to
native implementations. NumPy provides array programming capabilities
through optimized C routines~\cite{numpy2020}, while PyTorch relies
heavily on C++ and CUDA backends to deliver high-performance tensor
operations~\cite{paszke2019pytorch}.

At the same time, alternative language ecosystems have emerged. Julia was
designed to overcome the so-called ``two-language problem,'' in which
prototypes are written in high-level languages and later rewritten in
lower-level languages for performance. Through LLVM-based just-in-time
compilation, Julia aims to provide both developer productivity and
near-native execution speed~\cite{bezanson2017,lin2021}. Rust has gained
increasing attention in AI infrastructure because its ownership model
provides memory safety without garbage collection, making it attractive
for performance-critical libraries and tooling. Go has similarly found
widespread use in distributed systems, cloud infrastructure, and
model-serving environments due to its simplicity, concurrency primitives,
and deployment characteristics.

Despite these developments, existing evaluations typically measure the
performance of language ecosystems rather than the languages themselves.
When implementations rely on optimized libraries, the dominant
computational work is often executed in C, C++, Rust, or GPU kernels
regardless of the language visible to the user. As a result, there remains
a need for controlled studies that isolate the effect of the programming
language itself through from-scratch implementations of AI algorithms.

\subsection{AI Algorithms as Benchmark Workloads}
Benchmark selection is a critical factor in programming-language
evaluation because different computational patterns show different aspects
of language runtimes and compiler optimizations. To obtain a broad
coverage of artificial intelligence workloads, this study focuses on five
well-established algorithms that represent distinct AI paradigms.

$K$-means clustering remains one of the most widely used unsupervised
learning techniques and serves as a representative example of iterative
distance-based optimization~\cite{macqueen1967,lloyd1982}.
$K$-nearest neighbors ($k$-NN) provides a classic instance-based learning
method whose computational cost is dominated by distance calculations and
comparison operations~\cite{cover1967}. Multilayer perceptrons trained
through backpropagation represent the foundation of modern neural-network
learning and exercise dense numerical computation and gradient-based
optimization~\cite{rumelhart1986}. Genetic algorithms exemplify
evolutionary computation through population-based search and stochastic
optimization~\cite{holland1992}. Finally, Mamdani fuzzy inference systems
remain among the most influential approaches in fuzzy logic and rule-based
intelligent systems~\cite{mamdani1975}.

These algorithms were selected not because they represent the current
state of the art, but because their definitions are mature, reproducible,
and broadly understood. Together, these algorithms cover a broad range of
computational patterns, including floating-point intensive computation,
exhaustive search, iterative optimization, stochastic evolution, and
rule-based inference. This diversity allows an examination of whether
language performance remains consistent across different types of AI
workloads. To the best of the authors' knowledge, no previous study has
compared from-scratch
implementations of such a diverse set of AI algorithms across Python, C,
C++, Rust, Go, and Julia using a common experimental framework.

\section{Methodology}\label{sec:method}

The comparison rests on a single idea: if every language solves a
byte-for-byte identical problem, then any gap in time or memory is a
property of the language, not of the computation. The rest of this
section makes that idea concrete, describing the languages and
toolchain (Sect.~\ref{sec:langs}), the five algorithms and their
parameters (Sect.~\ref{sec:algos}), the determinism protocol that
equalizes the workload across languages (Sect.~\ref{sec:fairness}),
and the measurement protocol (Sect.~\ref{sec:metrics}).

\subsection{Languages and Toolchain}\label{sec:langs}
The six languages (Python, C, C++, Rust, Go, and Julia) were motivated in
Sect.~\ref{sec:related}; the concrete toolchain on which they run is fixed
here. They span the design space along three axes that matter for AI
workloads: execution model (interpreted, JIT-compiled, ahead-of-time
compiled), memory management (garbage-collected vs.\ manual/ownership),
and abstraction level. Table~\ref{tab:langs} lists the compiler or runtime
and the
optimization settings used for each. All experiments run on an Apple
MacBook with an M1 Pro processor (six performance and two efficiency
cores) and 16~GB of unified memory, under macOS. Each language uses its
standard, idiomatic optimization settings; none uses SIMD intrinsics,
hand-vectorization, or external numerical libraries.

\begin{table}[h]
\caption{The six languages, their toolchains, and optimization settings.
Versions are pinned in the reproducibility artifact.}
\label{tab:langs}
\centering
\begin{tabular*}{\linewidth}{@{\extracolsep{\fill}}lll}
\toprule
Language & Toolchain & Optimization \\
\midrule
Python & CPython 3.14.3 & interpreted (no flags) \\
C      & Apple Clang 17.0.0 & \texttt{-O3 -march=native} \\
C++    & Apple Clang 17.0.0 & \texttt{-O3 -march=native -std=c++17} \\
Rust   & rustc 1.93.0 & \texttt{opt-level=3}, \texttt{lto=true}, \texttt{codegen-units=1} \\
Go     & gc 1.26.3 & \texttt{go build} (default) \\
Julia  & 1.9.2 & JIT (default) \\
\bottomrule
\end{tabular*}
\end{table}

\subsection{Benchmark Algorithms}\label{sec:algos}
The five benchmarks are sized so that the fastest languages, C and C++, run
for roughly $0.35$--$0.75$\,s apiece: long enough to dwarf timer noise,
short enough that even Python finishes the full suite in hours rather than
days. Table~\ref{tab:params} lists the parameters; the descriptions below
fix the details that must match across languages for the outputs to agree:
initialization, tie-breaking, and stopping rules. Several of the choices
below (squared distances, ReLU with mean-squared error, the Rosenbrock
objective, and triangular membership functions) follow from the
bit-exactness requirement motivated in Sect.~\ref{sec:fairness}.

\begin{description}
  \item[$k$-means (data mining).] Lloyd's algorithm on $N{=}100{,}000$
    points in $D{=}4$ dimensions with $K{=}10$ clusters and $1000$ fixed
    iterations (no early stopping, for determinism). Centroids are
    initialized to the first $K$ points; each point is assigned to the
    nearest centroid by squared Euclidean distance, ties broken to the
    lowest index; centroids are recomputed as cluster means.
  \item[$k$-NN (machine learning).] $M{=}50{,}000$ labeled training
    points and $Q{=}10{,}000$ queries in $D{=}8$ dimensions over $3$
    classes. Each query is classified by majority vote over its
    $k{=}15$ nearest training points (squared Euclidean distance), with
    vote ties broken to the lowest class label.
  \item[MLP (neural networks).] A fully connected network
    $16{\rightarrow}64{\rightarrow}4$ trained by full-batch gradient
    descent for $150$ epochs at learning rate $0.01$ on $N{=}10{,}000$
    synthetic samples. The hidden layer uses ReLU activation and the
    output is linear under a mean-squared-error loss; weights are
    initialized in $[-0.1, 0.1]$ and biases at zero.
  \item[GA (computational intelligence).] A genetic algorithm minimizing
    the $D{=}30$ Rosenbrock function with a population of $5000$ over
    $1200$ generations: tournament selection (size $3$), uniform
    crossover, uniform mutation (rate $0.1$, half-range $0.1$), and
    elitism preserving the best individual. Genes are initialized
    uniformly in $[-5, 5]$.
  \item[Fuzzy (fuzzy systems).] A Mamdani inference system with two
    inputs, three triangular membership functions per input, a nine-rule
    base, max--min aggregation, and centroid defuzzification over a
    $100$-point output grid, evaluated on $2{\times}10^{6}$ input pairs.
\end{description}

\begin{table}[h]
\caption{The five benchmarks, one per branch of AI, with key
parameters.}
\label{tab:params}
\centering
\begin{tabular*}{\linewidth}{@{\extracolsep{\fill}}lll}
\toprule
Benchmark & AI area & Key parameters \\
\midrule
$k$-means & Data mining & $N{=}100{,}000$, $D{=}4$, $K{=}10$, $1000$ iterations \\
$k$-NN & Machine learning & $M{=}50{,}000$, $Q{=}10{,}000$, $D{=}8$, $k{=}15$, $3$ classes \\
MLP & Neural networks & $16{\rightarrow}64{\rightarrow}4$, $150$ epochs, ReLU, MSE \\
GA & Comp. Intell. & $D{=}30$, pop.\ $5000$, $1200$ gens, tournament-$3$ \\
Fuzzy & Fuzzy systems & $2$ inputs, $3$ MFs, $9$ rules, $2{\times}10^{6}$ inferences \\
\bottomrule
\end{tabular*}
\end{table}

\subsection{Cross-Language Fairness and Determinism}\label{sec:fairness}

Cross-language fairness is enforced through two design choices. First,
each language implements the same algorithm with identical parameters,
written idiomatically but without external libraries. Second, all
randomness is generated by a single 64-bit linear congruential
generator (LCG), reproduced verbatim in all six languages:

\begin{equation}
s_{i+1} = (a\,s_i + c) \bmod 2^{64},
\qquad
u_i = (s_i \gg 33)\,/\,2^{31},
\end{equation}

with multiplier $a = 6364136223846793005$, increment
$c = 1442695040888963407$, and seed $s_0 = 42$. Because both the
generator and the order in which its values are consumed are identical
across implementations, every benchmark receives the same sequence of
pseudo-random inputs.

To enable bit-for-bit output comparisons, computations are restricted
to basic IEEE-754 arithmetic operations standard~\cite{ieee754} (addition, subtraction,
multiplication, division, and min/max), minimizing
implementation-dependent numerical differences. Transcendental
functions (\texttt{exp}, \texttt{log}, \texttt{cos}, \texttt{sqrt})
are deliberately avoided because their implementations may differ
across language runtimes and math libraries; for example, Go ships its
own math library rather than relying on the system
\texttt{libm}. This constraint shapes each benchmark. Nearest-neighbour
search compares squared distances rather than invoking
\texttt{sqrt}; the neural network uses ReLU activations and
mean-squared error instead of sigmoid or softmax functions, which
require \texttt{exp}; the genetic algorithm optimizes the polynomial
Rosenbrock function rather than the cosine-based Rastrigin function and
uses uniform mutation instead of Gaussian noise; and the fuzzy system
employs triangular membership functions.

Each benchmark concludes by reporting two fingerprints of its output:
an integer fingerprint (for example, a count of cluster assignments or
a sum of predicted labels) and a floating-point summary statistic.
Across all six languages, integer fingerprints are identical, while
floating-point summaries agree to the six decimal places reported in
this paper. Integer fingerprints are particularly valuable because they
are unaffected by the one remaining source of low-level numerical
variation considered here: fused multiply--add (FMA) contraction,
which compilers may introduce under flags such as
\texttt{-march=native}.

Agreement of both fingerprints provides strong evidence that all
implementations perform equivalent computations. Under these
conditions, observed differences in execution time and memory
consumption can be attributed to language implementations, compilers,
and runtime systems rather than algorithmic divergence.

\subsection{Metrics and Measurement Protocol}\label{sec:metrics}
Five metrics are reported, covering both runtime cost and developer
cost.

\textbf{M1 (wall-clock time)} and \textbf{M2 (peak resident set size)}
are measured together. Each benchmark is executed under
\texttt{/usr/bin/time -l}, which reports \texttt{ru\_maxrss} on macOS,
and timed with \texttt{hyperfine} using 10 measured runs after 3 warmup
runs. Standard output is redirected to \texttt{/dev/null}, although each
program still computes and prints its checksum to prevent dead-code
elimination.

\textbf{M3 (binary size)} is the size of the compiled executable as
reported by \texttt{stat}. It is not applicable to Python and Julia.
\textbf{M4 (lines of code)} is measured with \texttt{wc -l}, summed
across the five benchmark implementations and excluding build files.
\textbf{M5 (compilation time)} is the cold-build wall time of the five
benchmarks after clearing the toolchain cache. Like binary size, it is
not applicable to Python and Julia.

To summarize performance across benchmarks, slowdown ratios are
aggregated with the geometric mean, the standard averaging method for
normalized measurements.

\subsection{Reproducibility}\label{sec:repro}
The study is deliberately not containerized. On Apple silicon, Docker
runs Linux inside a virtual machine, changing the operating system,
system libraries, and toolchains while introducing additional
virtualization overhead. Since the objective is to measure native
execution time and memory consumption on the target platform, all
experiments are performed directly on the host system.

Reproducibility instead relies on the determinism protocol described in
Sect.~\ref{sec:fairness}. Every implementation executes the same
algorithm, consumes the same pseudo-random inputs, and produces the same
validation fingerprints. As a result, any machine can reproduce the
workload exactly, although absolute execution times may differ across
hardware and software environments.

All source code, benchmark harnesses, raw result files, and toolchain
versions are included in the artifact. The complete benchmark suite can
be reproduced with a single command
(\texttt{bash harness/run\_all.sh}).

\section{Results}\label{sec:results}

The runtime metrics are presented first (M1 wall-clock time, M2 peak
memory), followed by the developer-cost metrics (M3 binary size, M4 lines
of code, M5 compilation time). All timings are the mean of ten runs after three
warmups; the relative standard deviation stayed under $4\%$ on every
language--benchmark pair and under $1.5\%$ in the large majority,
confirming a quiet measurement environment.

\subsection{Execution Time (M1)}
\begin{figure}[!htbp]
\centering
\includegraphics[width=\textwidth]{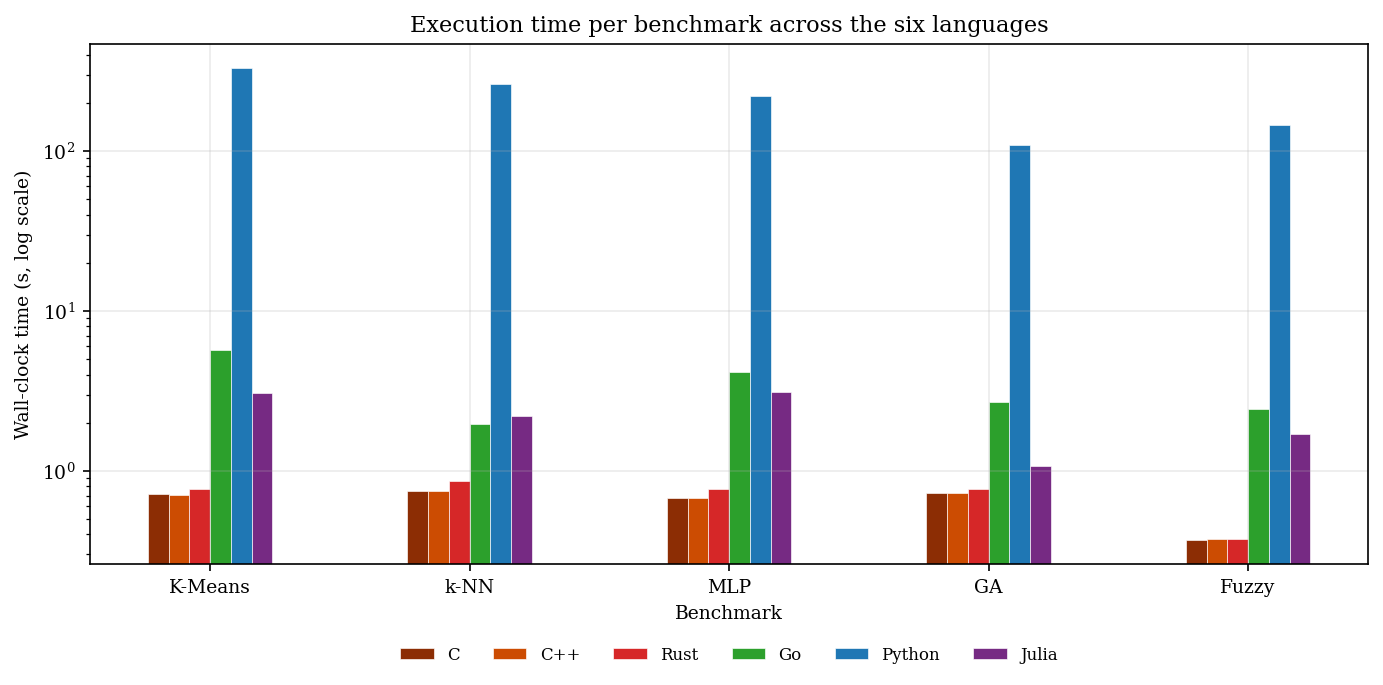}
\caption{Mean wall-clock time per benchmark and language (log scale).
The three compiled systems languages (C, C++, Rust) are visually
indistinguishable; Go and Julia occupy a middle band; Python sits two to
three orders of magnitude above.}\label{fig:walltime}
\end{figure}

Table~\ref{tab:time} reports mean wall-clock time for all thirty
language--benchmark pairs, together with each language's geometric-mean
slowdown relative to C. Figure~\ref{fig:walltime} shows the same data on a
logarithmic scale.

\begin{table}[!htbp]
\caption{Mean wall-clock time in seconds (ten runs after three warmups).
The last column is the geometric mean of each language's per-benchmark
slowdown relative to C. Lower is better; the fastest entry per benchmark
is in bold.}\label{tab:time}
\centering
\setlength{\tabcolsep}{9pt}
\renewcommand{\arraystretch}{1.2}
\begin{tabular}{lrrrrrr}
\toprule
Language & $k$-means & $k$-NN & MLP & GA & Fuzzy & Geo.\ $\times$C \\
\midrule
C      & 0.717 & 0.748 & 0.678 & 0.728 & \textbf{0.369} & 1.00 \\
C++    & \textbf{0.706} & \textbf{0.746} & \textbf{0.675} & \textbf{0.722} & 0.372 & 1.00 \\
Rust   & 0.769 & 0.861 & 0.774 & 0.772 & 0.375 & 1.09 \\
Julia  & 3.062 & 2.200 & 3.113 & 1.074 & 1.698 & 3.30 \\
Go     & 5.710 & 1.969 & 4.174 & 2.688 & 2.451 & 5.01 \\
Python & 332.0 & 261.2 & 220.4 & 108.7 & 144.7 & 314.6 \\
\bottomrule
\end{tabular}
\end{table}

Three tiers emerge. C and C++ are statistically indistinguishable: C++
wins four of five benchmarks by under $2\%$ and C wins the fifth (fuzzy)
by under $1\%$. Rust trails this pair by a geometric mean of $9\%$, with
its largest gap on $k$-NN ($+15\%$) where bounds-checked indexing into the
training set is on the hot path. Julia ($3.30\times$) and Go ($5.01\times$)
form a middle tier in which, against intuition, the JIT-compiled language
is \emph{faster} than the ahead-of-time-compiled one on four of five
benchmarks. Python forms a distant tier at $314.6\times$ C, ranging from
$149\times$ on the branch-heavy genetic algorithm to $463\times$ on the
arithmetic-dense $k$-means.

Notably, Go's slowdown is far from uniform, ranging from $2.6\times$ on
$k$-NN to $8.0\times$ on $k$-means. This workload dependence is the central
evidence for RQ3; its cause is analyzed in Sect.~\ref{sec:discussion}.

\FloatBarrier
\subsection{Peak Memory (M2)}
\begin{figure}[!htbp]
\centering
\includegraphics[width=\textwidth]{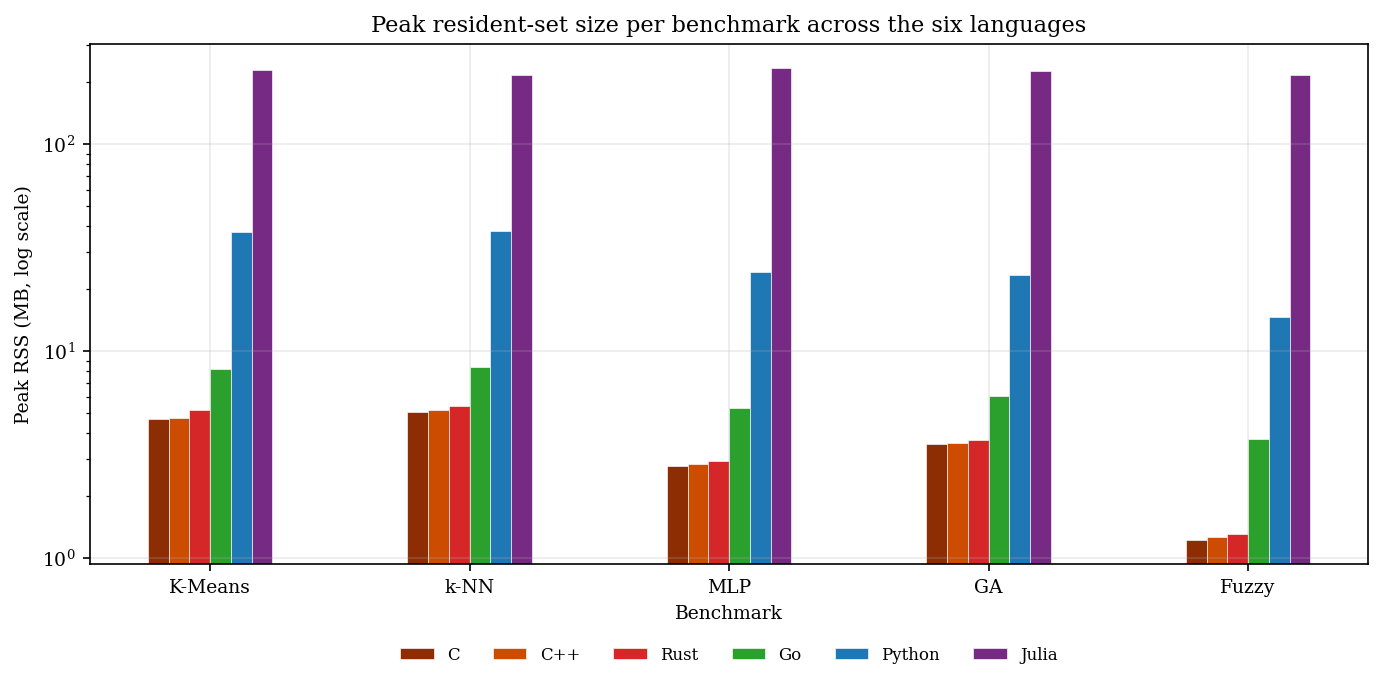}
\caption{Mean peak resident set size per benchmark and language (log
scale). Julia's runtime footprint is roughly two orders of magnitude
above the systems languages and is essentially constant across
workloads.}\label{fig:rss}
\end{figure}

Table~\ref{tab:mem} reports mean peak resident set size, visualized in
Fig.~\ref{fig:rss}. The manual- and ownership-managed languages (C, C++,
and Rust) remain below $6$\,MiB on every benchmark. Go's
garbage-collected runtime increases average memory usage to
$6.3$\,MiB. Python's interpreter overhead raises this figure to roughly
$28$\,MiB. Julia is the clear outlier: its JIT compiler and runtime
maintain approximately $\sim\!224$\,MiB resident regardless of the
workload, a fixed cost that dwarfs the benchmark data itself and may
become a significant consideration in memory-constrained deployments.

\begin{table}[!htbp]
\caption{Mean peak resident set size in MiB (\texttt{ru\_maxrss} via
\texttt{/usr/bin/time -l}). Lower is better; lowest per benchmark in
bold.}\label{tab:mem}
\centering
\setlength{\tabcolsep}{9pt}
\renewcommand{\arraystretch}{1.2}
\begin{tabular}{lrrrrrr}
\toprule
Language & $k$-means & $k$-NN & MLP & GA & Fuzzy & Mean \\
\midrule
C      & \textbf{4.7} & \textbf{5.1} & \textbf{2.8} & \textbf{3.6} & \textbf{1.2} & \textbf{3.5} \\
C++    & 4.8 & 5.2 & \textbf{2.8} & \textbf{3.6} & 1.3 & 3.5 \\
Rust   & 5.2 & 5.4 & 2.9 & 3.7 & 1.3 & 3.7 \\
Go     & 8.1 & 8.4 & 5.3 & 6.1 & 3.8 & 6.3 \\
Python & 37.6 & 38.1 & 24.2 & 23.2 & 14.7 & 27.6 \\
Julia  & 228.7 & 216.2 & 234.3 & 225.3 & 216.5 & 224.2 \\
\bottomrule
\end{tabular}
\end{table}

\FloatBarrier
\subsection{Developer-Cost Metrics (M3--M5)}
Table~\ref{tab:static} reports binary size, lines of code, and
compilation time. Binary size spans nearly two orders of magnitude:
C and C++ produce $\sim\!34$\,KB executables, Rust $373$\,KB
(due to monomorphization and a statically linked runtime), and Go
$2.5$\,MB (including its runtime and garbage collector).

Lines of code, summed across the five benchmark implementations, are
surprisingly similar across languages. C++ ($397$) and C ($406$) are
the most concise, while Python ($406$) requires essentially the same
amount of code as C. Go ($541$) is the most verbose, largely due to
explicit error handling and the use of conventional loops for numeric
kernels.

Compilation is sub-second per benchmark for C and C++. Rust's
$8.7$\,s cold build reflects its heavier optimization pipeline and
monomorphization, which contribute to its runtime performance.

\begin{table}[!htbp]
\caption{Developer-cost metrics. M3 (binary size) and M5 (compile time)
are undefined for the interpreted languages. M4 is summed across the
five benchmark implementations; M5 measures the cold-build time of the
complete compiled benchmark suite for each language.}
\label{tab:static}
\centering
\setlength{\tabcolsep}{14pt}
\renewcommand{\arraystretch}{1.2}
\begin{tabular}{lrrr}
\toprule
Language & M3 binary (KB) & M4 lines of code & M5 compile (s) \\
\midrule
C      &   33.6 & 406 & 0.93 \\
C++    &   33.7 & 397 & 1.13 \\
Rust   &  373.1 & 474 & 8.73 \\
Go     & 2509.4 & 541 & 3.30 \\
Python &    --- & 406 & --- \\
Julia  &    --- & 514 & --- \\
\bottomrule
\end{tabular}
\end{table}

\begin{figure}[!htbp]
\centering
\includegraphics[width=0.7\textwidth]{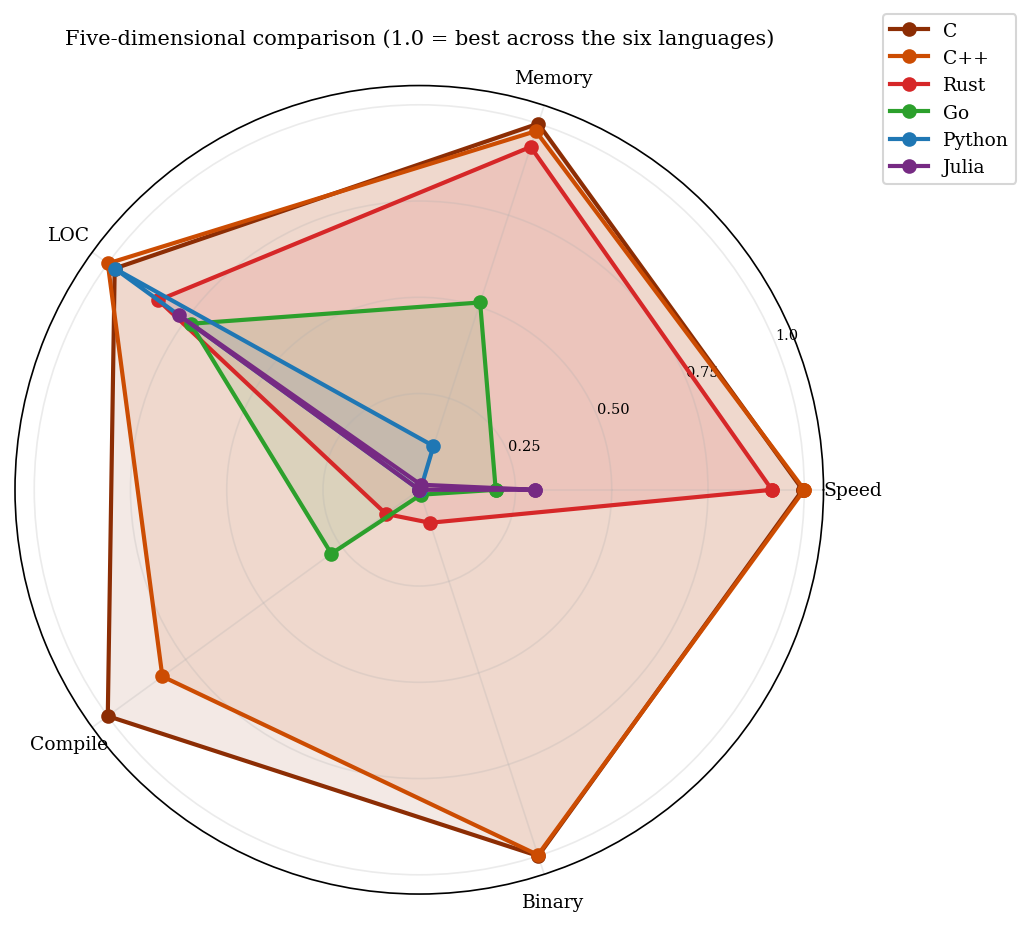}
\caption{Normalized five-metric profile of all six languages (each axis
scaled so that larger is better). The C and C++ outlines nearly coincide,
confirming their tie; the systems languages dominate every runtime axis,
while the interpreted and JIT languages recover ground only on developer
cost (LOC).}\label{fig:radar}
\end{figure}

Figure~\ref{fig:radar} consolidates all five metrics on a common
normalized scale, summarizing the overall profile of each language.

\FloatBarrier
\section{Discussion}\label{sec:discussion}

\subsection{RQ1: Time and Memory Ranking}

The three execution-time tiers identified in
Sect.~\ref{sec:results} provide a direct answer to RQ1, but the
implications become clearer when execution time and memory consumption
are considered together. C, C++, and Rust are observed to form a
high-performance tier whose execution times differ by less than
$10\%$ on average. Within this group, the choice is unlikely to be
driven by performance alone. Instead, considerations such as memory
safety, tooling, and ecosystem maturity become increasingly important.

A more substantial distinction appears in the managed-language tier.
Julia achieves considerably better execution times than Go, but this
advantage is accompanied by a much larger memory footprint. Across all
benchmarks, Julia maintains a resident memory usage of approximately
$224$,MiB regardless of workload size, whereas Go remains below
$10$,MiB. As a result, the relative attractiveness of these languages
depends strongly on deployment constraints. Where execution speed is
the primary concern, Julia may be preferable; where memory availability
is limited, Go presents a more balanced alternative.

More broadly, execution-time rankings and memory rankings do not
necessarily coincide. Language selection therefore depends not
only on raw performance but also on the resource constraints of the
target environment.

\subsection{RQ2: The Productivity--Performance Trade-off}

A common assumption in AI development is that higher-level languages
offer substantially greater productivity in exchange for reduced
performance. The results only partially support this view.

Perhaps the most surprising observation is that Python does not exhibit
a meaningful conciseness advantage in the from-scratch setting studied
here. The total lines of code required to implement the five benchmarks
are nearly identical to C and slightly higher than C++
(Table~\ref{tab:static}). Without external numerical libraries, the
algorithms must be expressed explicitly through loops, indexing, and
scalar operations, which largely removes Python's usual abstraction
advantages.

This observation helps explain a broader phenomenon in modern
AI software. Much of Python's practical productivity stems from access
to highly optimized libraries rather than from the execution
characteristics of the language itself. By intentionally excluding
external libraries, the study isolates the contribution of the language
and highlights the extent to which AI productivity often depends on the
compiled ecosystems beneath Python.

The trade-offs associated with Rust and Go are more conventional.
Rust incurs longer compilation times and produces larger binaries than
C and C++, yet delivers performance within $9\%$ of C while providing
memory-safety guarantees. Go produces the largest binaries and the most
verbose implementations, reflecting the cost of a managed runtime and
garbage collection. These results reinforce the idea that productivity,
safety, compilation cost, and runtime performance should be viewed as
complementary dimensions rather than independent metrics.

Rust's compilation cost deserves particular attention because it is
closely tied to the optimization strategy used in this study. The
configuration \texttt{lto=true} and \texttt{codegen-units=1} increases
build times but also contributes to Rust's strong runtime performance.
The $8.7$,s compilation time is therefore best viewed not as an isolated
drawback, but as part of a broader trade-off between development-time
and execution-time efficiency.

\subsection{RQ3: Does the Winner Depend on the Workload?}

The results indicate that language rankings are not completely stable
across AI workloads. Although C, C++, and Rust consistently occupy the
top tier, the relative performance of Go and Julia varies considerably
depending on the computational characteristics of the benchmark.

The clearest example is Go, whose slowdown relative to C ranges from
approximately $2.6\times$ on $k$-NN to $8.0\times$ on $k$-means.
This variation suggests that different workloads expose different
strengths and limitations of the underlying compiler and runtime.
Benchmarks such as $k$-means, MLP, and fuzzy inference spend most of
their execution time in dense floating-point computations, whereas
$k$-NN performs a larger proportion of comparisons and indexing
operations. The resulting performance differences are substantial
enough to influence language-selection decisions in practice.

Julia exhibits workload sensitivity as well, although in a different
manner. Its strongest result occurs on the genetic algorithm benchmark,
where the branch-heavy execution pattern reduces the advantage enjoyed
by aggressively optimized systems languages. In contrast, Julia falls
further behind C on workloads dominated by floating-point reductions.

Taken together, these findings provide a clear answer to RQ3: no single
language dominates across all categories of AI computation. The
performance gap between a well-matched language and a poorly matched
one can exceed an order of magnitude, suggesting that workload
characteristics should play a central role in implementation-language
selection.

\subsection{Threats to Validity}

Several limitations should be considered when interpreting these
results. First, all measurements were collected on a single hardware and
operating-system configuration (Apple M1 Pro running macOS). Although
the broad performance tiers are expected to generalize to other platforms,
their exact magnitudes may differ on x86-64 systems or under different
compiler toolchains.

Second, the study focuses on from-scratch implementations and excludes
external numerical libraries, SIMD intrinsics, and hand-tuned
optimizations. Consequently, the results characterize the languages and
their standard toolchains rather than the maximum performance achievable
through expert optimization.

Third, Julia's measurements include startup and JIT compilation costs.
These costs are relevant for short-lived processes and scripting
workloads but may be amortized in long-running applications.

Finally, the requirement of cross-language numerical reproducibility led
to the avoidance of transcendental functions whose implementations differ
across platforms. While this decision improves experimental control, it also
means that the benchmarks do not capture all computational patterns
present in modern AI workloads.

\section{Conclusion}\label{sec:conclusion}

This paper presented a controlled empirical comparison of six programming languages (Python, C, C++, Rust, Go, and Julia) through from-scratch implementations of five representative artificial intelligence algorithms spanning data mining, machine learning, neural networks, computational intelligence, and fuzzy systems. By enforcing identical algorithmic specifications, deterministic input generation, and cross-language output verification, the study isolated the effects of language runtimes, compilers, and memory-management models from those of external libraries and frameworks.

The results reveal three clear performance tiers. C and C++ consistently achieved the fastest execution times and lowest memory consumption, with Rust remaining within 9\% of their performance while exhibiting a similarly small memory footprint. Julia and Go formed an intermediate tier, although with markedly different trade-offs: Julia generally achieved faster execution times but incurred a substantially larger runtime memory overhead, whereas Go maintained moderate memory usage at the cost of slower execution on several workloads. Python was consistently the slowest language by a large margin, highlighting the extent to which its practical success in AI depends on the optimized native libraries that underpin its ecosystem.

When considering execution time, memory consumption, code size, and compilation cost together, the results suggest that language selection for AI systems cannot be reduced to a single metric. The observed trade-offs vary substantially across languages and, importantly, across workloads. While the relative ranking of C, C++, and Rust remained stable, the performance of Julia and Go showed significant sensitivity to algorithm characteristics, demonstrating that workload composition plays a major role in determining language efficiency.

Overall, the findings answer the three research questions by showing that: (1) substantial differences exist in execution time and memory consumption among the evaluated languages; (2) performance advantages are accompanied by trade-offs in compilation cost, binary size, and implementation characteristics; and (3) language rankings are not entirely consistent across AI workloads, indicating that the computational profile of an algorithm should be an important consideration when selecting an implementation language.

Future work will extend the benchmark suite to additional AI workloads, including modern deep-learning kernels and transformer-related operations, as well as evaluate the portability of the observed trends across operating systems, processor architectures, and hardware accelerators. Incorporating SIMD optimizations, GPU implementations, and widely used AI libraries would also help quantify the gap between intrinsic language performance and the performance achieved in real-world AI software stacks.

%
%
%
%

\end{document}